%% file: main.tex
\documentclass[10pt,conference]{IEEEtran}
\ifCLASSINFOpdf
\else
\fi
\input{preamble}

\input{paper-specific}

\usepackage{etoolbox}

\begin{document}
%
\title{Scalable Quantum Error Correction for Surface Codes using FPGA}


\author{\IEEEauthorblockN{Namitha Liyanage,
Yue Wu, Alexander Deters and
Lin Zhong}
\IEEEauthorblockA{Department of Computer Science,
Yale University, 
New Haven, CT\\
Email : \{namitha.liyanage, yue.wu, alex.deters, lin.zhong\}@yale.edu}}


%



\maketitle
\thispagestyle{plain}
\pagestyle{plain}

\begin{abstract}
A fault-tolerant quantum computer must decode and correct errors faster than they appear. The faster errors can be corrected, the more time the computer can do useful work. The Union-Find (UF) decoder is promising with an average time complexity slightly higher than $O(d^3)$. We report a distributed version of the UF decoder that exploits parallel computing resources for further speedup. Using an FPGA-based implementation, we empirically show that this distributed UF decoder has a sublinear average time complexity with regard to $d$,  given $O(d^3)$ parallel computing resources. The decoding time per measurement round decreases as $d$ increases, a first time for a quantum error decoder. The implementation employs a scalable architecture called Helios that organizes parallel computing resources into a hybrid tree-grid structure. We are able to implement $d$ up to 21 with a Xilinx VCU129 FPGA, for which an average decoding time is 11.5 ns per measurement round under phenomenological noise of 0.1\%, significantly faster than any existing decoder implementation. Since the decoding time per measurement round of Helios decreases with $d$, Helios can decode a surface code of arbitrarily large $d$ without a growing backlog.

\end{abstract}


%
\IEEEpeerreviewmaketitle

\input{introduction}

\input{background}
\input{design}

\input{system}

\input{implementation}

\input{results}

\input{related_work}

\input{conclusion}



\clearpage
\bibliographystyle{IEEEtran}
\balance
\bibliography{ref}{}

\end{document}

%% file: preamble.tex
\usepackage{setspace}
\usepackage{hyperref}
\usepackage{amsmath}
\usepackage[numbers, sort&compress]{natbib}

\usepackage{graphicx}
\usepackage{comment}

\newcommand{\be}{\begin{eqnarray}}
\newcommand{\ee}{\end{eqnarray}}

\usepackage{listings}
\usepackage[font=footnotesize]{caption}
\usepackage[font=footnotesize]{subcaption}
\usepackage{multirow} 

\usepackage{url}

\newcommand{\secref}[1]{\S\ref{#1}}

\newcommand{\lineref}[1]{L\ref{#1}}

\usepackage{enumitem}

%% file: paper-specific.tex
\usepackage{xcolor,soul}

\usepackage[ruled,linesnumbered]{algorithm3e}
\renewcommand{\algorithmautorefname}{Algorithm}
\usepackage{balance}

\newcommand{\code}[1]{\texttt{#1}}

\newcommand\couldremove[1]{{\st{#1}}}
\newcommand{\remove}[1]{}

\definecolor{lightgray}{gray}{0.6}
\definecolor{lightblue}{rgb}{0.9,0.9,1}
\definecolor{aqua}{rgb}{0.0, 1.0, 1.0}

\newcommand{\hlc}[2][aqua]{{\sethlcolor{#1}\hl{#2}}}
\newcommand\note[1]{\hlc[aqua]{#1}}
\newcommand\lin[1]{\hlc[yellow]{LZ: #1}} 
\newcommand\lina[1]{\hlc[gray]{LZ: #1}}
\newcommand\yue[1]{{\hlc[orange]{YW: #1}}} 
\newcommand\yuea[1]{\hlc[gray]{YW: #1}}
\newcommand\nami[1]{{\hlc[pink]{NL: #1}}}
\newcommand\namia[1]{{\hlc[gray]{NL: #1}}}

\newcommand\name[0]{Helios\xspace}

\newcommand\todo[1]{\textcolor{red}{TODO: #1}}

\renewcommand\couldremove[1]{}  
\renewcommand\note[1]{}         
\renewcommand\lin[1]{}          
\renewcommand\lina[1]{}         
\renewcommand\nami[1]{}          
\renewcommand\namia[1]{}         
\renewcommand\yue[1]{}          
\renewcommand\yuea[1]{}         
\renewcommand\todo[1]{}         


%% file: introduction.tex
\section{Introduction}
\label{sec:introduction}

The high error rates of quantum devices pose a significant obstacle to the realization of a practical quantum computer.  As a result, the development of effective quantum error correction (QEC) mechanisms is crucial for the successful implementation of a fault-tolerant quantum computer.

One promising approach for QEC is surface codes~\cite{dennis2002topological, fowler2012surface, bonilla2020xzzx} in which information of a single qubit (called a logical qubit) is redundantly encoded across many physical data qubits, with a set of ancillary qubits interacting with the data qubits.
By periodically measuring the ancillary qubits, one can detect and potentially correct errors in physical qubits.

Once the presence of errors has been detected through the measurement of ancillary qubits, a classical algorithm, or \emph{decoder}, guesses the underlying error pattern and corrects it accordingly.
The faster errors can be corrected, the more time a quantum computer can spend on useful work. 
Due to the error rate of the state-of-the-art qubits, very large surface codes ($d>25$) are necessary to achieve fault-tolerant quantum computing~\cite{fowler2012surface, Chen2021Exponential, Gidney2021rsa}. See \S\ref{sec:background} for more background. 

As surveyed in \S\ref{sec:related}, previously reported decoders capable of decoding errors as fast as measured, or \emph{backlog-free}, either exploit limited parallelism~\cite{das2021liliput, das2022afs, higgott2023sparse}, or sacrifice accuracy~\cite{Holmes2020Nisq, Ueno2021Qecool}. 
Sparse Blossom
~\cite{higgott2023sparse} and Fusion Blossom~\cite{yueGithub} feature an important algorithmic breakthrough in realizing MWPM-based decoders. Fusion Blossom can additionally leverage measurement round-level parallelism to meet the throughput requirement of very large $d$. However, due to their software-based realizations, both Sparse Blossom and Fusion Blossom suffer from decoding time per round longer than that of \name by orders of magnitude at large $d$ and higher noise level. When used in a quantum computer, the computer would spend most of execution time waiting for error correction.
\lin{Check}


In this paper we report a \textit{distributed Union-Find (UF) decoder} (\S\ref{sec:Design}) and its FPGA implementation called \textit{\name} (\S\ref{sec:System}). Given $O(d^3)$ parallel resources, our decoder achieves sublinear average time complexity according to empirical results for $d$ up to 21, the first to the best of our knowledge.
Notably,
adding more parallel resources will not reduce the time complexity of the decoder, due to the inherent nature of error patterns.
Our decoder is a distributed design of and logically equivalent to the UF decoder first proposed in~\cite{delfosse2017almost}. 
We implement the distributed UF decoder with \name, a scalable architecture for organizing the parallel computation units. 
\name is the first architecture of its kind that can scale to arbitrarily large surface codes by exploiting parallelism at the vertex level of the model graph. 
In \S\ref{sec:results}, we present experimental validations of the distributed UF decoder and \name using a VCU129 FPGA board~\cite{mpsoc} for up to $d=21$.
The decoder's average decoding time per measurement round under a phenomenological noise of 0.1\% is 11.5~ns for $d=21$, which is significantly faster than any existing decoder implementation.
Our results successfully demonstrate, for the first time, a decoder design with decreasing average time per measurement round when $d$ increases.
This shows evidence that the decoder can scale to arbitrarily large surface codes without a growing backlog.

In summary, we report the following contributions in this work. 
\begin{itemize}
    \item A distributed algorithm that implements the Union-Find decoder that can exploit parallel computing units to stop decoding time per measurement round from growing with the code distance $d$.

    \item The \name architecture and its FPGA-based implementation that realize the distributed Union-Find decoder. 

    \item A set of empirical data based on the FPGA implementation that demonstrate decreasing decoding time per round as $d$ grows and 11.5 ns decoding time per measurement round for $d=21$ under a phenomenological noise of 0.1\%.
\end{itemize}

\name is open-source and available from~\cite{qecGithub}.

%% file: background.tex
\section{Background}
\label{sec:background}

\input{background_sc.tex}

\input{background_decoding.tex}

%% file: background_sc.tex
\subsection{Error Correction and Surface Code}
\label{ssec:sc}

Quantum Error Correction (QEC) is more challenging than classical error correction due to the nature of Quantum bits. 
First, qubits cannot be copied to achieve redundancy due to the no-cloning theorem. 
Second, the value of the qubits cannot be directly measured as measurements perturb the state of qubits. 
Therefore QEC is achieved by encoding the \emph{logical state} of a qubit, as a highly entangled state of many physical qubits.
Such an encoded qubit is called a \emph{logical} qubit.

\begin{figure} [!t]
	\centering
 \begin{minipage}{.28\textwidth}
    \centering
	\begin{subfigure}{.98\textwidth}
	    \centering
        \includegraphics[width=1\textwidth]{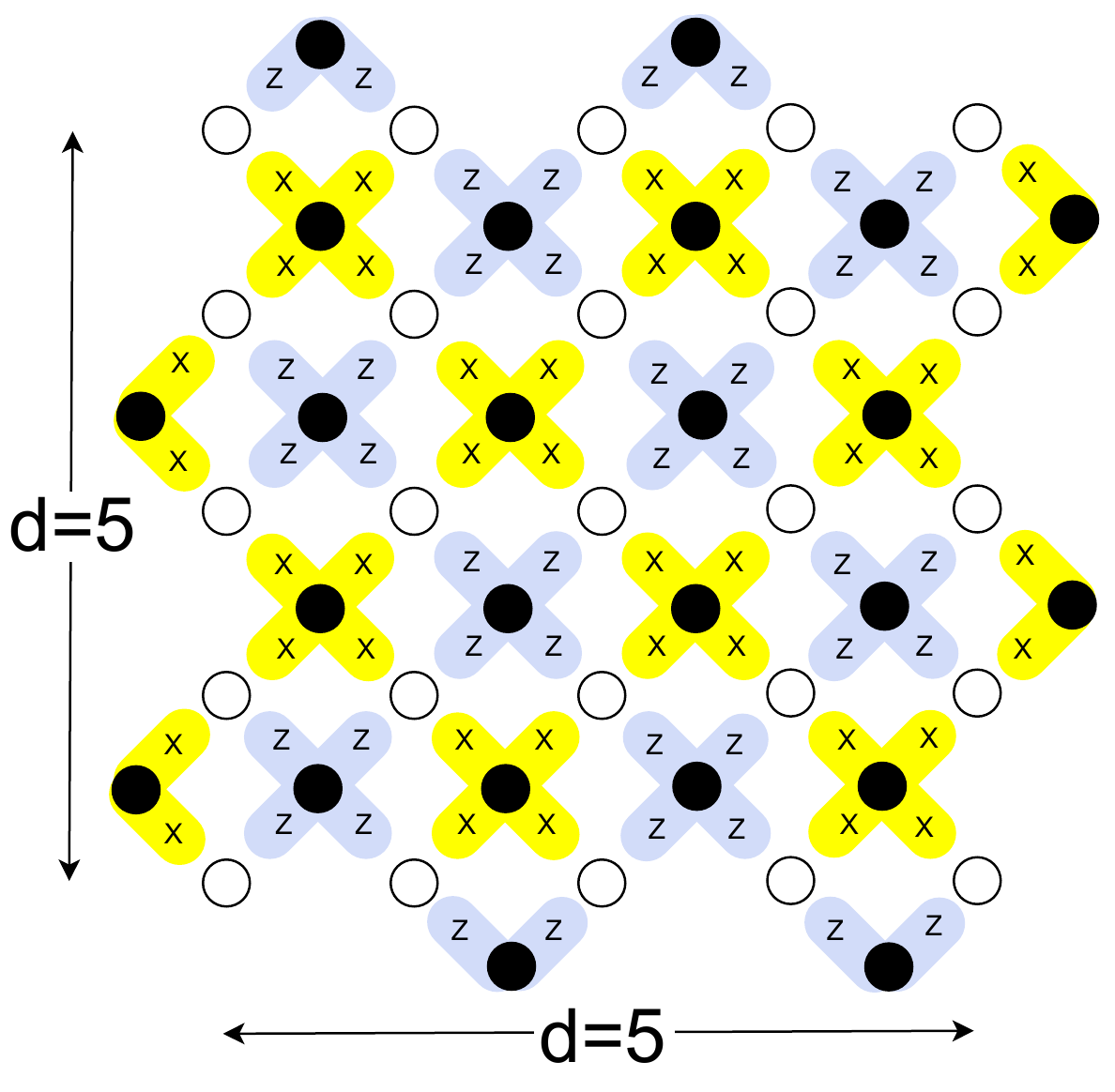}
        \caption{}
        \label{fig:basic_CSS}
    \end{subfigure}
\end{minipage}
 \begin{minipage}{.18\textwidth}
    \centering
	\begin{subfigure}{.98\textwidth}
	    \centering
        \includegraphics[width=1\textwidth]{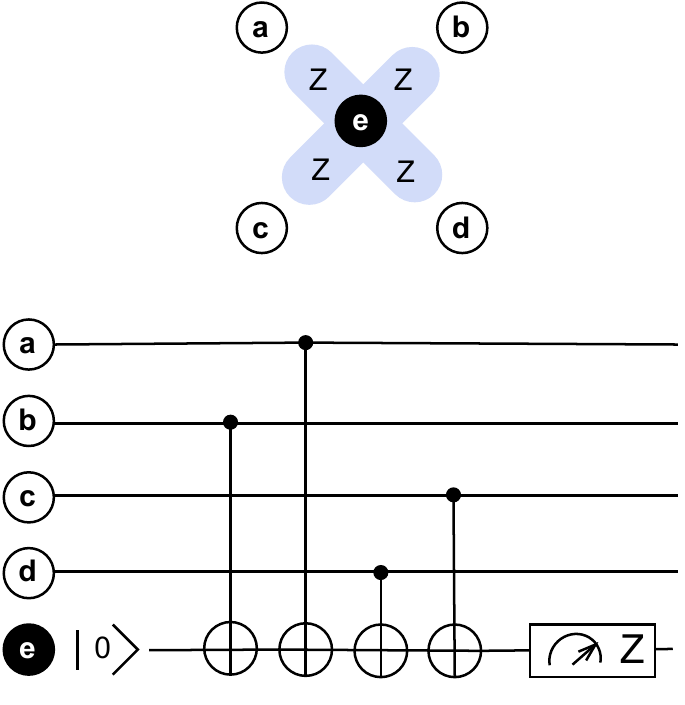}
        \caption{}
        \label{fig:CSS_Z_circuit}
    \end{subfigure}
    \hfill
	\begin{subfigure}{.98\textwidth}
	    \centering
        \includegraphics[width=1\textwidth]{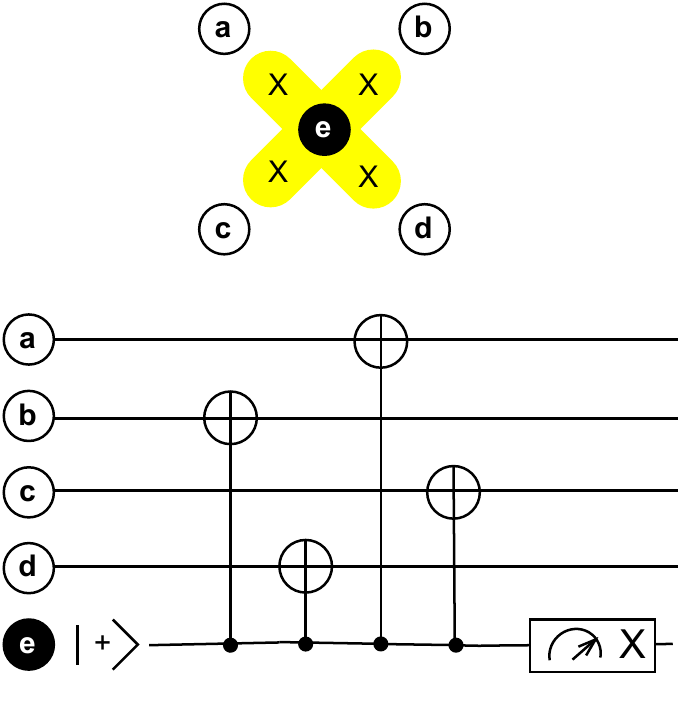}
        \caption{}
        \label{fig:CSS_X_circuit}
    \end{subfigure}
\end{minipage}
	\caption{(a) : Rotated CSS surface code ($d=5$), a commonly used type of surface code. The white circles are data qubits and the black are the Z-type and X-type ancillas. (b) and (c): Measurement circuit of Z-type and X-type ancillas. Excluding the ancillas in the border, each Z-type and X-type ancilla interacts with 4 adjacent data qubits.}
	\label{fig:surfacecode}
\end{figure}

The surface code is the widely used error correction code for quantum computing due to its high error correction capability and ease of implementation due to only requiring connectivity between adjacent qubits. 
A distance $d$ rotated surface code is a topological code made out of $2d^2 - 1$ physical qubits arranged as shown in \autoref{fig:surfacecode}.
A key feature of surface codes is that a larger $d$ can exponentially reduce the rate of logical errors making them advantageous.
For example, even if the physical error rate is 10 times below the threshold, $d$ should be greater than 17 to achieve a logical error rate below $10^{-10}$~\cite{fowler2012surface}.

\begin{figure} [!t]
	\centering
\centering
    \begin{minipage}{.30\textwidth}
    \centering
    \begin{subfigure}{0.98\textwidth}
	    \centering
        \includegraphics[width=1\textwidth]{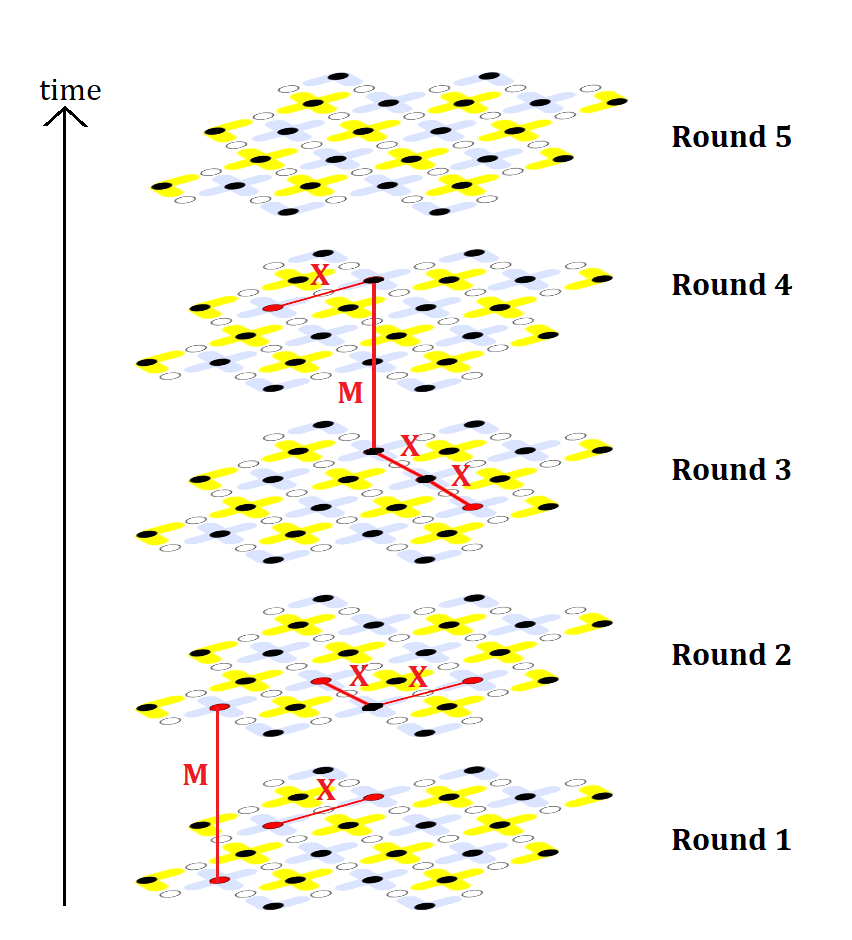}
        \caption{}
        \label{fig:example_syndrome}
        \end{subfigure}
        \end{minipage}
        \begin{minipage}{0.14\textwidth}
    \centering
        	\begin{subfigure}{0.98\textwidth}
	    \centering
        \includegraphics[width=\textwidth]{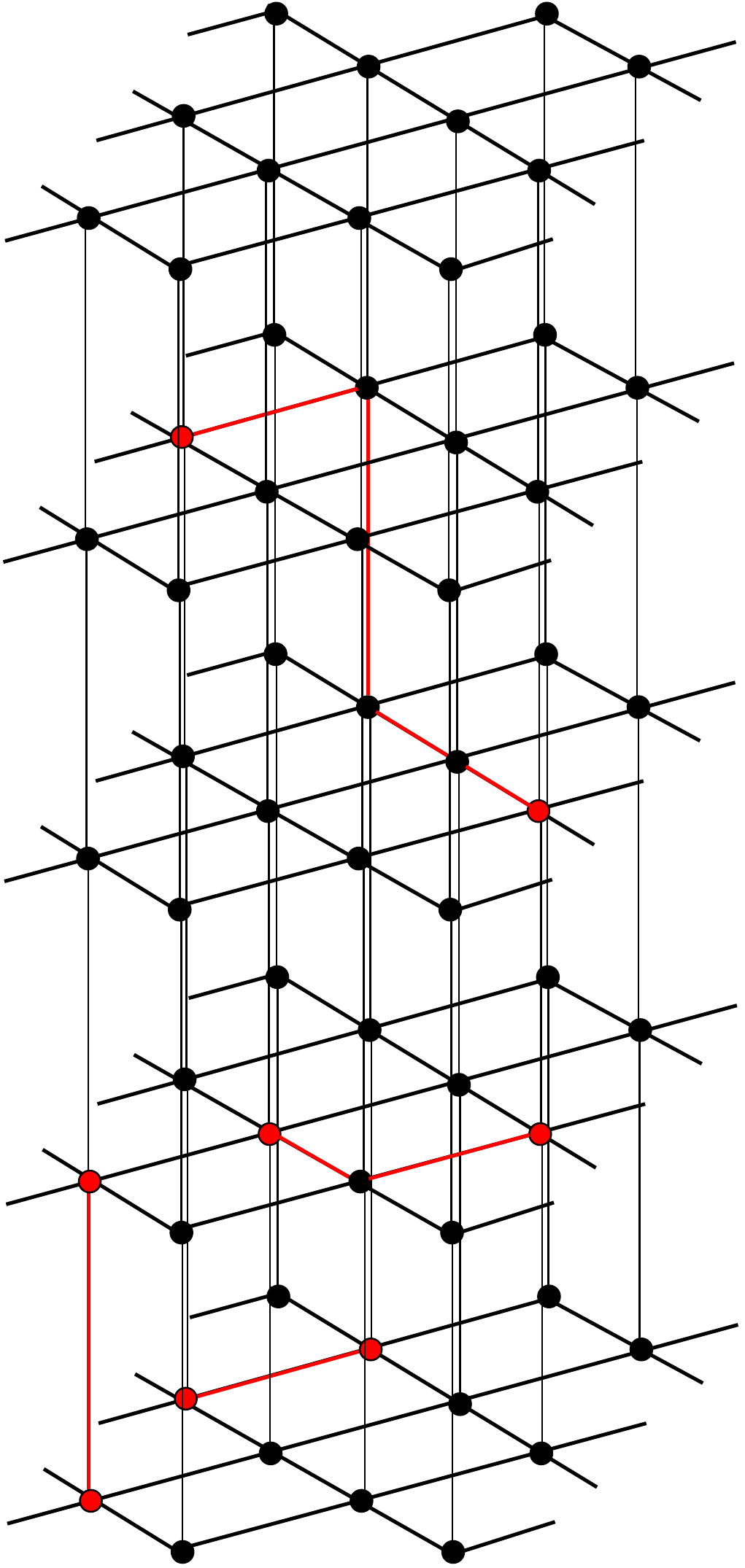}
        \caption{}
        \label{fig:decoding_graph}
    \end{subfigure}
    \end{minipage}
	\caption{(a) : An example syndrome of Z stabilizers for $d=5$ surface code with 5 rounds of measurements. The syndrome contains an isolated X-error (round 1), an isolated measurement error (rounds 1 and 2), a chain of two X errors (round 3), and a chain containing X errors and measurement errors spanning multiple measurement rounds (rounds 3 and 4). (b) Decoding graph with defect vertices marked red for the syndrome in (a). }
 
	\label{fig:error_patterns}
\end{figure}

A surface code contains two types of qubits, namely data qubits and ancilla qubits.
The data qubits collectively encode the \emph{logical state} of the qubit.
The ancilla qubits (called X-type and Z-type) entangle with the data qubits and by periodically measuring the ancilla qubits, physical errors in all qubits can be discovered and corrected.
An X error occurring in a data qubit will flip the measurement outcome of Z ancilla qubits connected with the data qubit and a Z error will flip the X ancilla qubits likewise. 
Such a measurement outcome is called \emph{defect measurement}.
Because  ancilla qubits themselves could also suffer from physical qubit errors, multiple rounds of measurements are necessary. 
The outcomes from these multiple rounds of measurements of ancilla qubits constitute a \emph{syndrome}.  
\autoref{fig:example_syndrome} shows a syndrome with sample physical qubit errors and shows how they are detected by ancilla qubits.
We only show X errors and measurement errors on Z-type ancillas because Z errors and measurement errors on X-type ancillas can be independently dealt with in the same way.

A syndrome can be conveniently represented by a graph
called \emph{decoding graph} in which a vertex represents a measurement outcome of an ancilla and an edge a data qubit. 
Vertices corresponding to defect measurements are specially marked.
The weight of an edge is determined by the probability of error in the corresponding data qubit or measurement.
 For distance $d$ surface code, there are $(d+1)\times (d-1)/2$ vertices.
This decoding graph can be extended to three-dimensional 
in which multiple identical planar layers are stacked on each other. 
Each layer represents a round of measurement. The minimum number of measurement rounds required to complete a fault-tolerant logical operation is $d$, which is also the number of rounds we consider in this paper.
Corresponding vertices in adjacent layers are connected by edges representing the corresponding ancilla's measurement error probability. 
That is, there are $(d+1)\times ((d-1)/2) \times d$ vertices in this three-dimensional graph. 
\autoref{fig:decoding_graph} shows the decoding graph for a syndrome from $d=5$ surface code.

%% file: background_decoding.tex
\subsection{Error Decoders}
\label{sec:decoder}
Given a syndrome, an error decoder identifies the underlying error pattern,  which will be used to generate a correction pattern. 
As multiple error patterns can generate the same syndrome, the decoder has to make a probabilistic guess of the underlying physical error.
The objective is that when the correction pattern is applied, the chance of the surface code entering a different logical state (i.e a logical error) will be minimized. 

\paragraph{Metrics} The two important aspects of decoders are accuracy and speed.
A decoder must correct errors faster than syndromes are produced to avoid a backlog. 
A faster decoder also allows more time for the quantum hardware to do actual useful work. 
The average decoding time per measurement round is a widely used criterion for speed.

A decoder must make a careful tradeoff between speed and accuracy.  A faster decoder with lower accuracy requires a larger $d$ to achieve any given logical error rate, which may require more computation overall.

 \paragraph{Union-Find (UF) Decoder} The UF decoder is a fast surface code decoder design first described by Delfosse and Nickerson~\cite{delfosse2017almost}. 
According to \cite{yue2021interpretation}, it can be viewed as an approximation to the blossom algorithm that solves minimum-weight perfect matching (MWPM) problems.
It has a worst-case time complexity of $O(d^3\alpha (d))$, where $\alpha$ is the inverse of Ackermann’s function, a slow-growing function that is less than three for any practical code distances. Based on our analysis, it has an average case time complexity slightly higher than $O(d^3)$.

\autoref{alg:serialuf} describes the UF decoder. It takes a decoding graph $\mathcal{G}(\mathbf{V},\mathbf{E})$ as input.
Each edge $e\in \mathbf{E}$ has a weight and a growth, denoted by $e.w$ and $e.g$, respectively. $e.g$ is initialized with $0$ and the decoder may grow $e.g$ until it reaches $e.w$. When that happens, we say the edge is \emph{fully grown}.

The decoder maintains a set of odd clusters, denoted by $\mathcal{L}$. $\mathcal{L}$ is initialized to include all $\{v\}$ that $v\in\mathbf{V}$ are defect measurements (\lineref{line:initial_list}). 
Each cluster $C$ keeps track of whether its cardinality is odd or even as well as its root element.

The UF decoder iterates over growing and merging the odd cluster list until there are no more odd clusters (inside the \textbf{while} loop of \autoref{alg:serialuf}). 
Each iteration has two stages: Growing and Merging. 
In the \emph{Growing} stage, each odd cluster ``grows'' by increasing the \emph{growth} of the edges incidental to its boundary. 
This process creates a set of \emph{fully grown} edges $\mathcal{F}$ (\lineref{line:grow-start} to \lineref{line:grow-end}). 
The Growing stage is the more time-consuming step as it requires traversing all the edges in the boundary of all the odd clusters and updating the global edge table. 
Since the number of edges is $O(d^3)$, the UF decoder is not scalable for surface codes with large $d$. 

In the \emph{Merging} stage, the decoder goes through each fully-grown edge to merge the two clusters connected by the edge using UNION($u,v$) operation.
The UNION($u,v$) merges the two clusters containing $u$ and $v$ by assigning a 
common root element to the two clusters.
When two clusters merge, the new cluster may become even. 

When there is no more odd cluster, the decoder finds a correction within each cluster and combines them to produce the correction pattern (\lineref{line:peeling}).

\begin{algorithm} [!t]
\DontPrintSemicolon
\caption{Union Find Decoder}\label{alg:serialuf}
\footnotesize
\setcounter{AlgoLine}{0}
\SetKwInOut{Input}{input}
\SetKwInOut{Output}{output}
\Input{A decoding graph $\mathcal{G}(\mathbf{V}, \mathbf{E})$ with X (or Z) syndrome}
\Output{A correction pattern}
\% Initialization\\
    \ForEach{$v\in \mathbf{V}$}{
        \If{$v$ is defect measurement}{
            Create a cluster $\{v\}$
        }
        \label{line:initial_list}
    }

\While{there is an odd cluster}{ \label{line:loop_start}
    \% Growing\\
    $\mathcal{F}\gets \emptyset$\\
    \ForEach{odd cluster $C$}{
    \label{line:grow-start}
         \ForEach{$e=<u,v>$, $u\in C, v\not\in C$ }{
            \If{$e.growth < e.w$}{ 
                $e.growth\gets e.growth+1$ \label{line:grow_serial}\\
                \If{$e.growth = e.w$}{ 
                    $\mathcal{F} \gets \mathcal{F} \cup \{e\}$ \label{line:fusion_set}
                }
            }
        }
        
    }\label{line:grow-end}
    \% Merging\\
    \ForEach{$e = <u,v> \in \mathcal{F}$}{
        UNION($u$, $v$) \label{line:merge}

    }
}
\label{line:loop_end}
\textup{Build correction within each cluster by constructing a spanning tree} \label{line:peeling}
\end{algorithm}

%% file: design.tex
\section{Distributed UF Decoder Design}
\label{sec:Design}

Our goal to build a QEC decoder is scalability to the number of qubits.
As surface codes can exponentially reduce logical error rate with respect to $d$, larger surface codes with hundreds or even thousands of qubits are necessary for fault-tolerant quantum computing. Therefore, the average decoding time per measurement round  should not grow with $d$, to avoid exponential backlog for any larger $d$.

We choose the UF decoder for two reasons.
First, it has a much lower time complexity than the MWPM algorithm. Although in general, the UF decoder achieves lower decoding accuracy than MWPM decoders, it is as accurate in many interesting surface codes and noise models~\cite{yue2021interpretation,huang2020fault}.
Second, the UF decoder maintains fewer intermediate states, which makes it easier to implement in a distributed manner.
We observe that the Growing stage from \lineref{line:grow-start} to \lineref{line:grow-end} in \autoref{alg:serialuf} operates on each vertex independently without dependencies from other vertices. 
A vertex requires only the parity of the cluster it is a part of for the growing stage.
Second, during the merging stage, a vertex only needs to interact with its immediate neighbors (\lineref{line:merge}).

\input{dalgorithm}

\begin{figure*} [!t]
	 \begin{minipage}{0.48\textwidth}   
     \centering
        \includegraphics[width=0.72\linewidth]{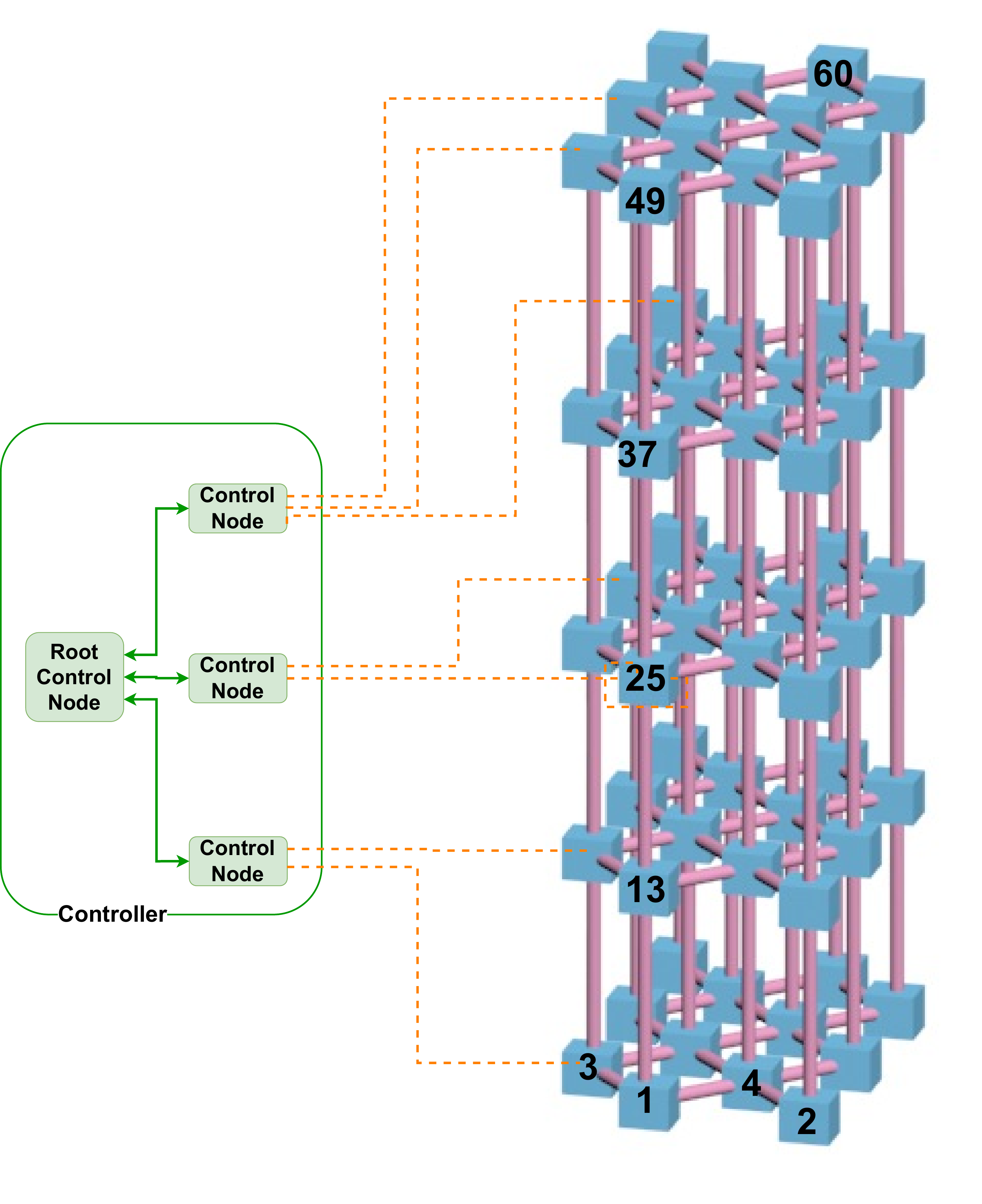}
	\caption{\name architecture for d=5 surface code for 5 measurement rounds. As d=5 surface code has 12 ancilla qubits of Z-type, \name contains a 12x5 PE array. PE $n$ indicates PE with $v.id=n$. Not all links from the controller to PEs and  all $v.id$s shown in the figure } 
 
 	\label{fig:pe_array}
\end{minipage}
\hfill
\begin{minipage}{0.48\textwidth}
    	    \centering
        \includegraphics[width=0.9\linewidth]{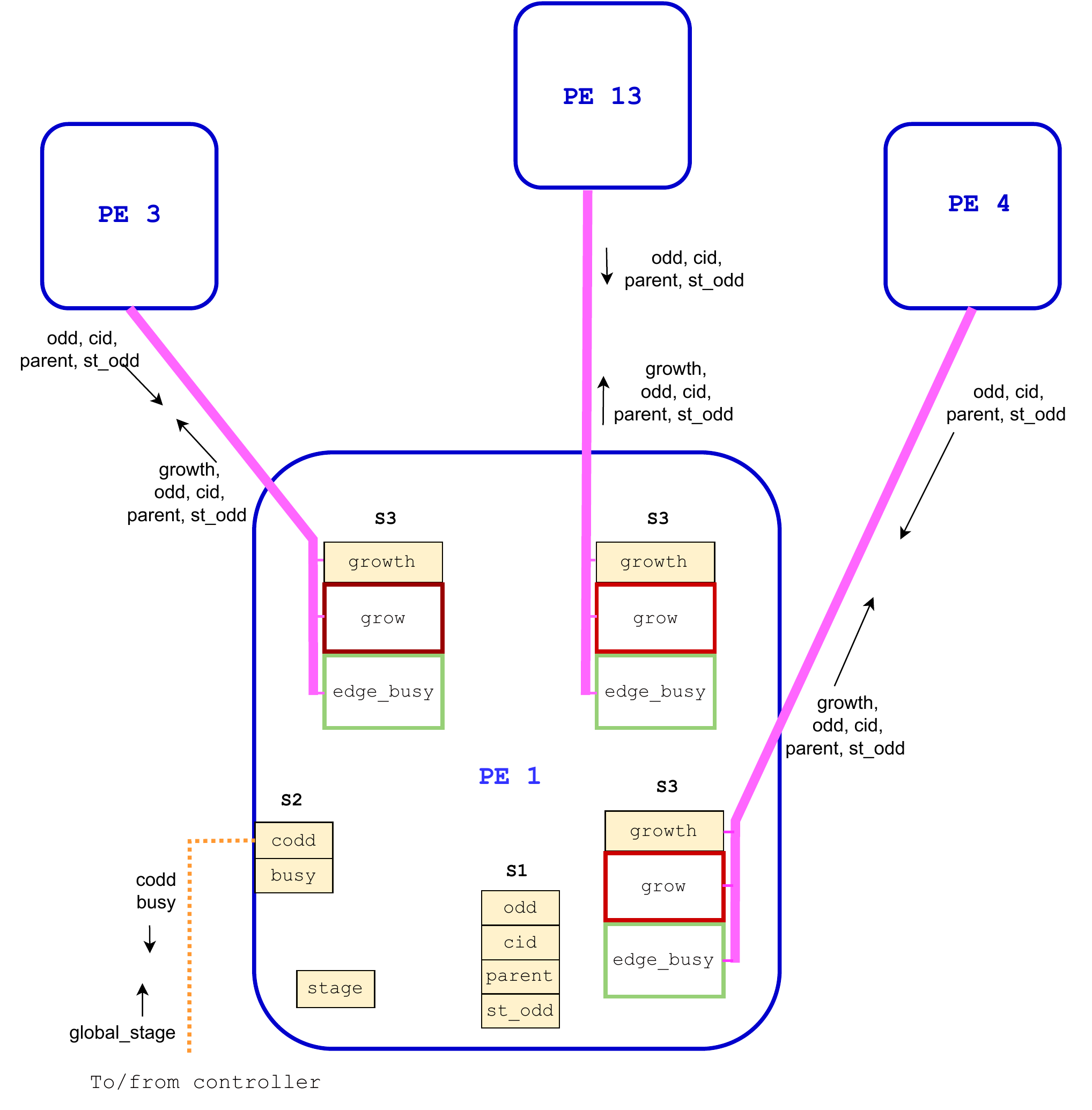}

\caption{The bottom left corner of the PE array shown in \autoref{fig:pe_array}. Only part of the logic and memory inside PE 1 is shown: \code{growth} (S3) is per edge and is stored in the PE with lower $id$. \code{grow} logic (in brown) calculates the updated growth value. \code{edge\_busy} (in green) is per adjacent PE and is used to calculate $v.$\code{busy}.
} 
  \label{fig:incidentPEs}
\end{minipage}
\end{figure*}

\subsection{Time Complexity Analysis}

We first show the PE coordination complexity and then calculate the overall time complexity based on that.

\paragraph{PE Coordination Complexity} The controller's time complexity is contingent upon the implementation of the shared memory for $v.\code{busy}$ and $v.\code{codd}$. 
Since both checks involve logical OR operations on individual PE information, the most efficient implementation consists of a logical tree of OR operations, yielding a time complexity of $O(log(d))$. 

\paragraph{Worst-case Time Complexity} The worst-case time complexity of our distributed UF decoder is $O(d^3log(d))$.
We explain this as follows.
Each stage of our distributed-UF algorithm is $O(1)$ time. Thus the worst case depends on the total number of stages. 
In the merging stage, both propagating the $cid$ and calculating the parity uses shared memory-based flooding and convergecast algorithms, each of which requires $O(D)$ merging and checking stages, where $D$ is the cluster diameter.
The maximum possible diameter, $O(d^3)$, occurs when a series of single-vertex clusters merge, creating a chain of clusters with a total diameter of $O(d^3)$.

As coordinating between stages has a complexity of $O(log(d))$, the overall time complexity is $O(d^3log(d))$.

Nevertheless, the worst-case scenario is extremely rare since larger clusters are exponentially less likely to occur. As shown in the empirical results reported in \secref{sec:results}, the average time grows sublinearly with $d$.

%% file: dalgorithm.tex
\subsection{Overview}
Like the original UF decoder, our distributed UF decoder is also based on the decoding graph.
Logically, the distributed decoder associates a processing element (PE) with each vertex in the graph. Therefore, when describing the distributed decoder, we often use PE and vertex in an inter-exchangeable manner. 
All PEs run the same algorithm, specified by \autoref{alg:distributed_uf}.
Like the UF decoder, a PE iterates over the \emph{Growing} and \emph{Merging} stages with the \emph{Merging} split into two: \emph{Merging} and \emph{Checking}.
Within each stage, PEs operate independently. 
A central controller coordinates their transition from one stage to the next as specified by \autoref{alg:distributed_uf_gc}.

A key challenge to the PE algorithm is to (\textit{i}) merge clusters and (\textit{ii}) compute the cluster parity, \emph{without} central coordination.
To achieve (\textit{i}), each PE is assigned a unique identifier (a natural number) and maintains the identifier of the cluster it belongs to, $cid$. The $cid$ is the lowest identifier of all its PEs. And the PE of the lowest identifier is called the root of the cluster. When two PEs connected by a fully grown edge have different $cid$s, the PE with the higher $cid$ adopts the lower value, resulting in the merging of their clusters. 
To achieve (\textit{ii}), each PE maintains a parent.
When a PE adopts the $cid$ from an adjacent PE, it sets the latter as its parent.
The parenthood relation between PEs creates a spanning tree for each cluster that is maintained by PEs locally and in which every PE in the cluster has a directional path to the root of the cluster.
The cluster parity can be computed using a convergecast algorithm on the spanning tree. 
We describe the PE algorithm in detail in~\ref{sec:pe_algorithm}.


To implement our distributed UF algorithm, we require several PE states, some of which are located in shared memories. We limit all communication between PEs and between PEs and the controller to coherent shared memories to ensure fast communication and prevent stalling that could result from message-based communication. 

\subsection{PE States}
A PE has direct read access to its local states and some states of incident PEs.
A PE can only modify its local states.

Thanks to the decoding graph, a PE has immediate access to the following objects.

\begin{itemize}[topsep=0.3em,itemsep=0.02em, leftmargin=*]
    \item $v$, the vertex it is associated with.
    
    \item $v.E$, the set of edges incident to $v$.

    \item $v.U$, the set of vertices that are incident to any $e\in v.E$ other than $v$ itself. We say these vertices are adjacent to $v$.
\end{itemize}

The algorithm augments the data structures of each vertex and edge of the decoding graph, according to the UF decoder design~\cite{delfosse2017almost}. 
For each vertex $v\in V$, the following information is added

 \begin{itemize}[topsep=0.3em,itemsep=0.02em, leftmargin=*]

 \item  $id$ : a unique identity number which ranges from $1$ to $n$ where $n=|V|$. $id$ is statically assigned and never changes. 

 \item  $m$ is a binary state indicating whether the measurement outcome is a defect measurement (\code{true}) or not (\code{false}). $m$ is initialized according to the syndrome.

\item  $cid$: a unique integer identifier for the cluster to which $v$ belongs, and is equal to the lowest $id$ of all the vertices inside the cluster. The vertex with this lowest $id$ is called the cluster root. $cid$ is initialized to be $id$. That is, each vertex starts with its own single-vertex cluster. 
When $cid=id$, the vertex is a root of a cluster. 

\item $odd$ is a binary state indicating whether the cluster is odd. $odd$ is initialized to be $m$.

\item $codd$ is a copy of $odd$.

\item $\code{parent}$ is a reference to the parent. As noted before, this parenthood relationship creates a spanning tree that  connects all vertices (PEs) with directional edges.

\item $st\_odd$: a binary state representing the parity of $m$ of $v$ and all its descendants.

\item $\code{stage}$ indicates the stage the PE currently operates in 

\item $\code{busy}$ is a binary state indicating whether the PE has any pending operations.

\end{itemize}

\begin{algorithm}[!t]
\DontPrintSemicolon
\caption{Algorithm for vertex $v$ in the distributed UF decoder.}\label{alg:distributed_uf}
\setcounter{AlgoLine}{25}
\footnotesize

$v.cid \gets v.id$;
$v.odd  \gets v.m$;
$v.parent \gets v.id$;
$v.st\_odd \gets v.m$ \;
\While{true}{
\If{$\code{global\_stage} = $\emph{\code{terminate}}}{
 \Return
}
    \textup{Wait until $\code{global\_stage} = ${\code{growing}}} \\
    growing($v$) \label{line:duf_grow_start} \\
    \textup{Wait until $\code{global\_stage} = ${\code{merging}}} \\
    \Do{$\code{global\_stage} = $\emph{\code{merging}} \label{line:merge_again}}
    { \label{line:do_in_PE}
    merging($v$) \\
    \textup{Wait until $\code{global\_stage} = ${\code{checking}}} \\
    checking($v$) \\
    \textup{Wait until $\code{global\_stage} != ${\code{checking}}}
    }
}
\end{algorithm}

\begin{algorithm}[!t]
\DontPrintSemicolon
\caption{Vertex growing algorithm}\label{alg:growing_function}
\footnotesize
\setcounter{AlgoLine}{40}
\function{growing(vertex v)}{
    $v.\code{busy} \gets \code{true}$; 
    $v.\code{stage} \gets \code{growing}$ \label{line:grow_response}\\
    \If{$v.odd$}{ \label{line:duf_is_odd}
         \ForEachAtomic{\textup{$e=\langle u,v\rangle\in v.E$}}{
                \If{$e.$\emph{\code{growth}}$<e.w$ \textup{and} $u.cid\neq v.cid$}{ \label{line:comparenupdate} 
                    $e.$\code{growth}$\gets e.$\code{growth}$+1$ \label{line:grow}
                }
        }
    }
    $v.\code{busy} \gets \code{false}$; \\
}

\end{algorithm}

\begin{algorithm} [!t]
\DontPrintSemicolon
\caption{Vertex merging algorithm}\label{alg:merging_function}
\setcounter{AlgoLine}{51}
\footnotesize

\function{merging(vertex v)}{
$v.\code{busy} \gets \code{true}$; 
$v.\code{stage} \gets \code{merging}$ \label{line:merge_response}\\
~\\


        \ForEach{$u\in v.nb$}{
            \If{$u.cid<v.cid$}{ \label{line:check_cid}
                $v.cid \gets u.cid$  \label{line:set_cid}\\
                $v.parent \gets u.id$ \label{line:set_parent}
            }
        }
        ~\\
        $v.st\_odd \gets \text{XOR}(u.st\_odd| u \in v.\code{child}, m)$ \label{line:subtree_parity}\\
        ~\\
        
        \lIf{$v.parent = v.id$}{
            $v.odd \gets v.st\_odd$ \label{line:root_parity}
        }
        \lElse{
            \textup{$v.odd \gets u.odd$ where $v.parent = u.id$} \label{line:calc_odd}
        }
        ~\\
        $v.\code{busy} \gets \code{false}$
}
\end{algorithm}

\begin{algorithm} [!t]
\DontPrintSemicolon
\caption{Vertex checking algorithm}\label{alg:checking_function}
\setcounter{AlgoLine}{68}
\footnotesize

\function{checking(vertex v)}{
$v.\code{busy} \gets \code{true}$\\
~\\

\If{\textup{$ \forall u \in v.\code{nb}, (u.cid = v.cid$ \& $v.odd = u.odd)$ and $v.st\_odd = \text{XOR}(w.st\_odd| w \in v.\code{child}, m)$ and $(v.parent \neq  v.id$ or $v.odd = v.st\_odd )$ }} { \label{line:merge_busy_check}
            $v.\code{busy} \gets \code{false}$
        }

$v.\code{stage} \gets \code{checking}$ \label{line:check_response} \\
}

\end{algorithm}

\begin{algorithm} [!t]
\DontPrintSemicolon 
\setcounter{AlgoLine}{76}
\footnotesize
\caption{The controller coordinates all PEs along stages and detects the presence of odd clusters.}\label{alg:distributed_uf_gc}
\While{true}{
    $\code{global\_stage} \gets \code{growing}$ \label{line:grow_signal}\\
    Wait until $\forall v \in V, v.\code{stage} = \code{growing}$\\
    Wait until $\forall v \in V, v.\code{busy} = \code{false}$

    ~\\
    \Do{\textup{$\exists v \in V, v.busy = \code{true}$}}{
    $\code{global\_stage} \gets \code{merging}$ \label{line:merge_signal}\\
    Wait until $\forall v \in V, v.\code{stage} = \code{merging}$\\
    Wait until $\forall v \in V, v.\code{busy} = \code{false}$ \\

    ~\\
    $\code{global\_stage} \gets \code{checking}$ \label{line:check_signal}\\
    Wait until $\forall v \in V, v.\code{stage} = \code{checking}$\\
    }

    ~\\
    \If{ \textup{$\forall v \in V, v.codd = \code{false}$} \label{line:check_odd}}{
        $\code{global\_stage} \gets \code{terminate}$ \label{line:terminate_signal}\\
        \Return\\
    }
}
\end{algorithm}

\noindent For each edge $e\in E$, the decoder maintains $e.$\code{growth}, which indicates the growth of the edge, in addition to $e.w$, the weight. $e.$\code{growth} is initialized as $0$. The decoder grows $e.$\code{growth} until it reaches $e.w$ and $e$ becomes \emph{fully grown}.

For clarity of exposition, we introduce a mathematical shorthand 
$v.$\code{nb}, the set of vertices connected with $v$ by full-grown edges, i.e., $v.$\code{nb}=$\{u|e=\langle v,u\rangle\in v.E~\land~e.$\code{growth}$=e.w\}$. We call these vertices the \emph{neighbors} of $v$. Note neighbors are always adjacent but not all adjacent vertices are neighbors.
We also use $v.$\code{child}, to indicate all child vertices of a vertex in the tree representation, i.e., $v.$\code{child}=$\{u|u.\code{parent} = v.id\}$. Since trees are built within a cluster, all child vertices are neighbors but not all neighbors are child vertices.

\subsection{Shared memory based communication}

We use coherent shared memory for a shared state that has a single writer.  For all shared memories, given the coherence, a read always returns the most recently written value. 
Like ordinary memory, we also assume both read and write are atomic.
\autoref{fig:incidentPEs} illustrates these memory blocks.

\begin{itemize} [topsep=0.3em,itemsep=0.02em, leftmargin=*]
    \item memory read/write for PE ($v$) and read-only for adjacent PEs, i.e., $\forall u\in v.U$. $v.id$, $v.cid$, $v.odd$, $v.parent$ and $v.st\_odd$ reside in this memory (S1).
    
    \item  memory read/write for PE ($v$) and read-only for the controller. 
    The PE local states,  $v.codd$, $v.$\code{stage} and $v.$\code{busy} reside in this memory (S2). 
    
    \item memory for  $e.$\code{growth}, which can be written by its two incident PEs (S3). 
    
    \item memory read/write for the controller and read-only for all PEs. The controller state \code{global\_stage} is stored in this memory (S4).
    
\end{itemize}

\subsection{PE Algorithm}
\label{sec:pe_algorithm}
All PEs iterate over three stages of operation. Within each stage, they operate independently but transit from one stage to the next when the controller updates \code{global\_stage}.
When a PE enters a stage, it sets $v.stage$ accordingly and keeps $v.\code{busy}$ as \code{true} until it finishes all work in the stage.
The controller uses these two pieces of information from all PEs to determine if a stage has started and completed, respectively (See \S\ref{sec:controller}).

We next describe the three stages of the PE algorithm.
In the \textbf{Growing} stage, vertices at the boundary of an odd cluster increase $e.$\code{growth} for boundary edges (\lineref{line:grow}).
As PEs perform Growing simultaneously, two adjacent PEs may compare $e.w$ and $e.growth$ and update $e.growth$ for the same $e$. 
Such compare-and-update operations must be atomic to avoid data race. 

In the \textbf{Merging} stage, two clusters connected through a fully-grown edge merge by adopting the lower cluster id ($cid$) of theirs.
To achieve this, each PE compares its $cid$ with its neighbors (\lineref{line:check_cid}).
If the other incident vertex of a fully grown edge has a lower $cid$, the PE adopts the lower $cid$ as its own (\lineref{line:set_cid}). 
The merging process continues until every PE in the cluster has the same $cid$, which is the lowest vertex identifier of the cluster.

In order to compute the cluster parity, when a PE adopts the $cid$ of the adjacent PE, it sets the latter as its \code{parent} (\lineref{line:set_parent}). 
This parenthood relation creates a spanning tree for each cluster that includes all PEs (vertices) with directional edges. 
Each PE then calculates the parity of itself and all its children as $st\_odd$ (\lineref{line:calc_odd}). 
Note that $odd$ of the root PE is the same as its $st\_odd$ (\lineref{line:root_parity}). 
All other PEs copy the $odd$ of their respective parents (\lineref{line:calc_odd}).

Astute readers may point out that $v.st\_odd$ should be the parity of $v$ and all its descendants, not just children.
This is achieved by two modifications, compared to the UF decoder.
First, a new stage \textbf{Checking} is added after Merging to see if the PE (vertex) needs to go back to \emph{Merging} again (\lineref{line:merge_busy_check}).
Second, all PEs iterates through Merging and Checking until all PEs have nothing to do for Merging. (\lineref{line:do_in_PE}-\lineref{line:merge_again}).
These allow parity computation to propagate from leaves to the roots of the spanning trees while $cid$ and $odd$ to propagate from the roots to the leaves. 

\paragraph{Building corrections within clusters}
While the original UF decoder builds a spanning tree within each even cluster in the end to generate a correction (\lineref{line:peeling}), 
our distributed UF decoder already has a spanning tree based on the parenthood relation and therefore is more efficient in generating corrections. 

\paragraph{Alternative Message-based Design}
Early on we considered the use of message-based communication to update the parity of a cluster~\cite{liyanage2023scalable}. 
This design requires directional links between PEs, with each PE serving as a router for forwarding messages, thus increasing the complexity of PEs.
Moreover, the finite capacity of directional links could lead to congested links, causing PE stalling, which in turn slowed down the decoding process and increased tail latency.

\subsection{Controller Algorithm}
\label{sec:controller}


The controller moves all PEs and itself along the three stages. 
In the Growing and Merging stages, it checks for $v.\code{busy}$ signals from each PE. 
The controller determines the completion of a stage when all PEs have $v.\code{busy}$ as \code{false}.
In the Checking stage controller determines the completion of the stage when all PEs have moved to the Checking stage.
Upon completion, the controller updates the \code{global\_stage} variable to move to the next stage and the PEs acknowledge this update by updating their own $v.\code{stage}$ variable.


The controller also calculates the presence of odd clusters. 
At the end of the Merging and Checking stages, it reads the $v.odd$ value of each vertex (\lineref{line:check_odd}). 
If any vertex has $v.odd = true$, the controller updates the global stage variable to Growing to continue the algorithm. 
Otherwise, it updates it to Terminate to end the algorithm.

%% file: system.tex
\section{\name Architecture}
\label{sec:System}

We next describe \name, the architecture for the distributed UF decoder.
\subsection{Overview}

\name organizes PEs and the controller in a custom topology that combines a 3-D grid and a tree as illustrated by \autoref{fig:pe_array} and explained below.
\begin{itemize}[topsep=0.3em,itemsep=0.02em, leftmargin=*]
    \item PEs are organized according to the position of vertices in the model graph they represent. We assign $v.id$ sequentially, starting with 1 from the bottom left corner and continuing in row-major order for each measurement round. 
    Shared memory S1 ($v.cid$, $v.odd$, $v.parent$ and $v.st\_odd$) and S2 ($v.codd$, $v.$\code{stage}, and $v.$\code{busy}) are per PE. 

    \item Shared memory S3 ($e.\code{growth}$) is added to the incident PE with the lower $id$. 
    
    \item A link between every two adjacent PEs to read from each other's S1 and for the one with the higher $id$ to read the other's S4. This results in a network of links in a 3-D grid topology. As a PE represents a vertex in the model graph, a link represents an edge. Broad pink lines in \autoref{fig:pe_array} represent these links.

    \item The controller is realized as a tree of control nodes (\secref{ssec:coordination_tree}). 
    The leaf nodes of the tree contain shared memory S4.

    \item A link between each PE and the controller for the controller to read from S2 and for the PEs to read from S4. Dashed orange lines in \autoref{fig:pe_array} represent these links.

\end{itemize}

\subsection{Controller}
\label{ssec:coordination_tree}
\name implements the controller as a tree of control nodes to avoid the scalability bottleneck. 
The controller requires three pieces of information from each PE: $v.codd$, $v.\code{stage}$ and $v.\code{busy}$.
Each leaf control node of the tree is directly connected with a subset of PEs. 
We can consider these PEs as the children of the leaf node.
Each node in the tree gathers vertex information from its children and reports it to the parent.
With information from all vertices, the root control node runs ~\autoref{alg:distributed_uf_gc} and decides whether to advance the stage. 

We leave height, branching factor, and the subset of PEs connected to each leaf node as implementation choices. 
The necessary requirement is that the controller should not slow down the overall design.






%% file: implementation.tex
\section{FPGA Implementation}
\label{sec:implementation}

We next describe an implementation of \name targeting a single FPGA.
We choose FPGA for two reasons. It supports massively parallel logic, which is essential as the number of PEs grows proportional to $d^3$ in our distributed UF design.
Moreover, it allows deterministic latency for each operation, which facilitates synchronizing all the PEs.
Our implementation contains approximately 3000 lines of Verilog code, which is publicly available at~\cite{qecGithub}.

\subsection{Leveraging global synchronization in FPGA}

We leverage global synchronization inside the FPGA to speed up our distributed UF algorithm. Running the FPGA design in a single-clock domain allows us to have all the PEs and the control nodes tightly synchronized. 
Notably, we simplify our algorithm as follows. 
Firstly, we run the Merging (\lineref{line:FPGA_merging}) and Checking stages (\lineref{line:FPGA_checking}) in parallel within each PE.
The tight synchronization of all PEs guarantees that false negative \code{busy} signals do not occur.

Secondly, we reduce the overhead of synchronization by having the controller only coordinate moving to the Growing stage at the beginning of each iteration (\lineref{line:PE_move_growing}). As each PE can perform the Growing stage deterministically in a single cycle, PEs can move to the Merging stage without central coordination (\lineref{line:PE_move_merging}). 

Additionally, as the controller deterministically knows the exact stage each PE is in, \code{stage} is stored locally and not shared with the controller. Thus the information from the PEs to the controller is limited to two bits, $v.\code{busy}$ and $v.\code{odd}$. 

\autoref{alg:FPGA_distributed_uf} and \autoref{alg:FPGA_gc} lists the FPGA-oriented algorithm of PE and the controller. The logic at every positive edge is executed in parallel. \autoref{fig:incidentPEs} shows a minimal diagram of a PE in the FPGA implementation.

\IncMargin{0.4em}
\begin{algorithm}[!t]
\setcounter{AlgoLine}{95}
\DontPrintSemicolon
\caption{FPGA-oriented algorithm for vertex $v$ in the distributed UF decoder.}\label{alg:FPGA_distributed_uf}
\footnotesize

$v.cid \gets v.id$;
$v.odd  \gets v.m$;
$v.parent \gets v.id$;
$v.st\_odd \gets v.m$ \;
~\\
\% Stage transition logic\\
\Posedge{}{
\lIf{$\code{global\_stage} = $\emph{\code{terminate}}}{
 \Return
}
\lElseIf{$\code{global\_stage} = $\emph{\code{growing}}}{
    $v.\code{stage} \gets \code{growing}$ \label{line:PE_move_growing} 
}
\lElseIf{$v.\code{stage} = $\emph{\code{growing}}}{
    $v.\code{stage} \gets \code{merging}$
    \label{line:PE_move_merging} 
}
}
~\\
\% Growing logic\\
\Posedge{}{
    \If{$v.\code{stage} = $\emph{\code{growing}}}{
         \ForEach{\textup{$e=\langle u,v\rangle\in v.E$ and $v.id < u.id$}}{
                \If{$e.$\emph{\code{growth}}$<e.w$ \textup{and} $u.cid\neq v.cid$}{ \label{line:FPGA_comparenupdate} 
                    \If{v.odd \textup{and} u.odd}{
                        $e.$\code{growth}$\gets \text{MIN}(e.$\code{growth}$+2$, $w$)
                    }
                    \ElseIf{v.odd \textup{or} u.odd}{
                        $e.$\code{growth}$\gets \text{MIN}(e.$\code{growth}$+1$, $w$)
                    }
                }
        }
    }
}
~\\
\% Merging logic \label{line:FPGA_merging}\\
\Posedge{}{
    Let $u$ be $\arg\min_{u \in (v.nb~\cup~\{v\})}(u.cid)$\\
    \If{$u.cid<v.cid$}{ 
        $v.cid \gets u.cid$  \\
        $v.parent \gets u.id$ 
    }
}
\Posedge{}{
$v.st\_odd \gets subtree\_parity(v)$
}
 \Posedge{}{       
    \lIf{$v.parent = v.id$}{
            $v.odd \gets v.st\_odd$ 
    }
    \lElse{
        \textup{$v.odd \gets u.odd$ where $u.id = v.parent$} 
    }
}
~\\
\% Checking logic \label{line:FPGA_checking}\\
\Posedge{}{
    \If{\textup{$ \exists u \in v.\code{nb}, (u.cid \neq v.cid\ \|\ v.odd \neq u.odd)$}}{
        $v.\code{busy} \gets \code{true}$\\
    }
    \ElseIf{\textup{ $v.st\_odd \neq subtree\_parity(v)$}}{
        $v.\code{busy} \gets \code{true}$\\
    }
    \ElseIf{\textup{ $(v.parent =  v.id$ \& $v.odd \neq v.st\_odd )$}}{
        $v.\code{busy} \gets \code{true}$\\
    }
    \Else{$v.\code{busy} \gets \code{false}$}
}
~\\
\function {subtree\_parity($v$)}{
$parity \gets v.m$\\
    \ForEach{$u \in v.\code{child}$}{
        $parity \gets \text{XOR}(parity, u.st\_odd)$
    }
\Return parity
}

\end{algorithm}

\begin{algorithm} [!t]
\DontPrintSemicolon 
\setcounter{AlgoLine}{160}
\footnotesize
\caption{FPGA-oriented controller logic}\label{alg:FPGA_gc}
$\code{global\_stage} \gets \code{growing}$ \\
\Posedge{}{
    \If{\textup{$\code{global\_stage} = \code{growing}$}} {
        $\code{global\_stage} \gets \code{merging}$ \\
        \%Wait until all PEs are in Merging Stage \\
        Wait 2 clock cycles
    }
    \ElseIf{ \textup{$\forall v \in V, v.\code{busy} = \code{false}$}} {
        \If{ \textup{$\forall v \in V, v.codd = \code{false}$}} {
            $\code{global\_stage} \gets \code{terminate}$ \label{line:FPGA_terminate}\\
        }
        \Else{
            $\code{global\_stage} \gets \code{growing}$
        }
    }
}
\end{algorithm}
\DecMargin{0.4em}

\paragraph{Time Complexity}
The worst-case time complexity of the FPGA design is $O(d^3)$ in contrast to $O(d^3log(d))$ of the generic distributed UF algorithm.
The $log(d)$ factor in the latter originates from the coordination overhead associated with transitioning between Merging and Checking stages. 
However, in the case of FPGA implementation, these two stages—Merging and Checking—are performed concurrently, obviating stage transitions. 
This concurrent operation effectively removes the $log(d)$ component.

\subsection{Implementation details}

We next list the other implementation choices of our design.

\textit{Controller}:~~
Since we only use a single FPGA and evaluate with $d$ up to 21, a single node controller suffices.
The node controller reads \code{busy} of each PE, every clock cycle to identify the completion of a stage.

\begin{figure}[t]
    \centering
        \begin{subfigure}{0.49\linewidth}
	    \centering
        \includegraphics[width=1\textwidth]{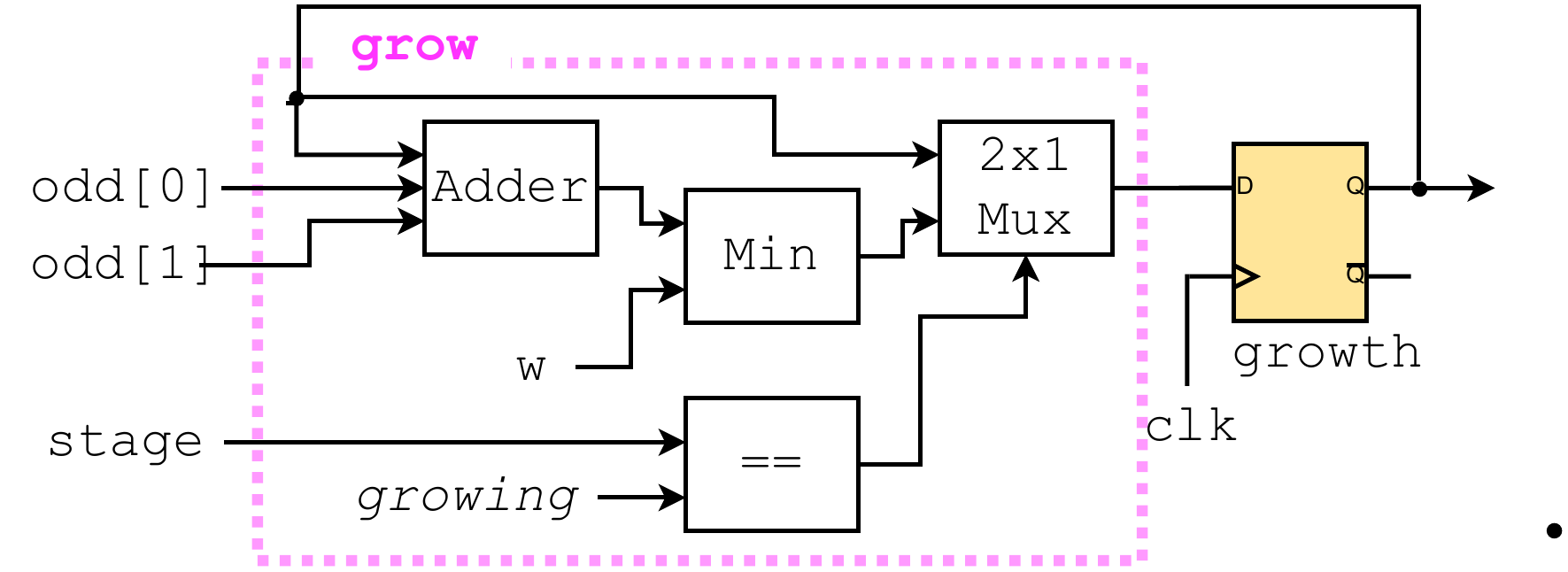}
        \label{}
        \end{subfigure}
    \begin{subfigure}{0.49\linewidth}
	    \centering
\begin{lstlisting}[language=Verilog, frame=single, basicstyle=\scriptsize\ttfamily]
reg growth;
always@(posedge clk)
  if(stage == growing)
    growth <= `MIN(growth 
      + odd[0] + odd[1], w);

\end{lstlisting}
\label{}
        \end{subfigure}
\caption{Circuit diagram of \code{grow} sub-module and Verilog implementation. This implements the  atomic compare and update operation in \lineref{line:comparenupdate} as part of the PE module. $odd[0]$ and $odd[1]$ represents the $odd$ state of the two incident PEs of the edge.}

\label{list:egrowth}
\end{figure}

\textit{Shared memory}:~~
We implement all shared memories as FPGA registers, i.e., \textbf{\code{reg}} in Verilog. 
FPGA registers by design guarantee that a read returns the last written value. 
In order to ensure that the S4 memory has a single writer, we adjust the PE logic to update growth by implementing a modified compare-and-update operation (\lineref{line:FPGA_comparenupdate}) as shown in~\autoref{list:egrowth}. The PE that houses the S3 memory performs this operation, increasing $e.\code{growth}$ by two when both endpoints of the edge have $v.odd$ set to true.

\subsection{Resource Usage}

On the VCU129 FPGA development board~\cite{vcu129}, we are able to support the distributed UF decoder with $d$ up to 21, due to resource limits. \autoref{tab:FPGA_usage} shows the resource usage for various $d$. 
While the numbers of vertices and edges grow by $O(d^3)$, resource usage grows faster for the following reasons. First, resource usage by a PE grows due to the increase of bit-width required for $v.id$, and $v.cid$. A PE for $d=21$ with six adjacent PEs requires 200 LUTs and a similar PE for $d=5$ requires only 155 LUTs.
Second, PEs on the surface of the three-dimensional array as shown in \autoref{fig:pe_array} use fewer resources than those inside because the latter have more incident edges. When $d$ increases a higher portion of PEs are inside the array.

We find that LUTs are the most critical resource in the FPGA for our design. 
It may be possible to run a design with $d=29$ on a Xilinx VU19 FPGA~\cite{vu19fpga}, which currently has the highest number of LUTs among commercially available FPGAs at the time of this writing.
Potentially larger $d$ values can be supported by using a network of FPGAs.

Existing commercial FPGAs like VCU129 often dedicate a lot of silicon to digital signal processing (DSP) units and block RAMs (BRAMs). 
However, our design does not use any DSPs because it only requires comparison operators and fixed point additions.
Our design does not use any BRAMs because all communication between PEs is shared memory based, which is implemented using registers.
Therefore, an ideal FPGA designed to run our distributed UF decoder would be simpler than current large FPGAs, as it would only need a large number of LUTs, no DSP units, and a limited amount of BRAM.

\begin{table}[t]
\caption{Resource usage of \name on VCU129 FPGA board for selected $d$}
\label{tab:FPGA_usage}
\small
\begin{center}
\begin{tabular}{|c|r|r|} \hline
$d$ & \# of LUTs & \# of registers \\
 \hline\hline
3        &  970    &  528          \\ \hline
5        &  6425    &   2425     \\ \hline
9        &  52111    &   13754      \\ \hline
13        &  165718    &   47211      \\ \hline
17        &  448314 & 122028 \\ \hline
21        &  898715 & 238939 \\ \hline
\end{tabular}
\end{center}
\end{table}

%% file: results.tex
\section{Evaluation}
\label{sec:results}

The main objective of our evaluation is to assess the scalability of our distributed UF implementation. To that end, we first describe our methodology and then show that the latency of our implementation grows sub-linearly with respect to the surface code size $d$. 

In addition, we also evaluate the impact of noise and non-identically distributed errors on latency.


\subsection{Methodology}
For speed, we measure the number of cycles required to decode a syndrome. 
To evaluate correctness, we compare the results of our distributed UF decoder with the results from the original UF decoder. 
We compare clusters because the original UF decoder and ours only differ in the clustering process.
In the rest of our evaluation, we will focus only on the speed of the distributed UF decoder and not on the accuracy of its results.

\paragraph{Experimental Setup}
As our evaluation setup, we use Xilinx VCU129 FPGA development board~\cite{mpsoc}, which is capable of decoding surface codes with $d$ up to 21. 

We use a MicroBlaze soft processor core~\cite{microblaze} instantiated inside the FPGA to generate the syndromes and transmit them to Helios, which runs on the same FPGA. We ran $10^6$ trials for each error rate and distance. 

\paragraph{Noise Model}
We use the phenomenological noise model~\cite{dennis2002topological} that accounts for errors in both data and ancilla qubits. 
As decoding for X-errors and Z-errors are independent and identical, we only focus on decoding X-errors in the evaluation.

To emulate noise, we independently flip the two adjacent stabilizer measurements for each data qubit with a probability of $p$ (the physical error rate) in each measurement round, and we also independently flip each stabilizer measurement with a probability of $p$ except for the first and last measurement rounds.
This is a widely used approach by prior QEC decoders ~\cite{das2022afs, Holmes2020Nisq, Skoric2022Parallel}. 
We then generate the syndrome from the physical errors and provide it as input to our decoder.

For most of our experiments, we use as default $p=0.001$, like other works~\cite{das2022afs}. 
This value is reasonable for surface codes, as $p$ should be sufficiently below the threshold (at least ten times lower) to exponentially reduce errors.
We note that the UF decoder has a threshold of $p=0.024$, calculated by Delfosse and Nickerson~\cite{delfosse2017almost}.


\subsection{Decoding Time}
\label{sec:decodingtime}
We experimentally show how the average time for decoding grows with the size of the surface code. 
Additionally, we show the effect of noise on the average time. 

\paragraph{Average time}

\begin{figure*}[!t]
        
	    \centering
        \begin{subfigure}{0.4\linewidth}
	    \centering
        \includegraphics[width=1.0\textwidth]{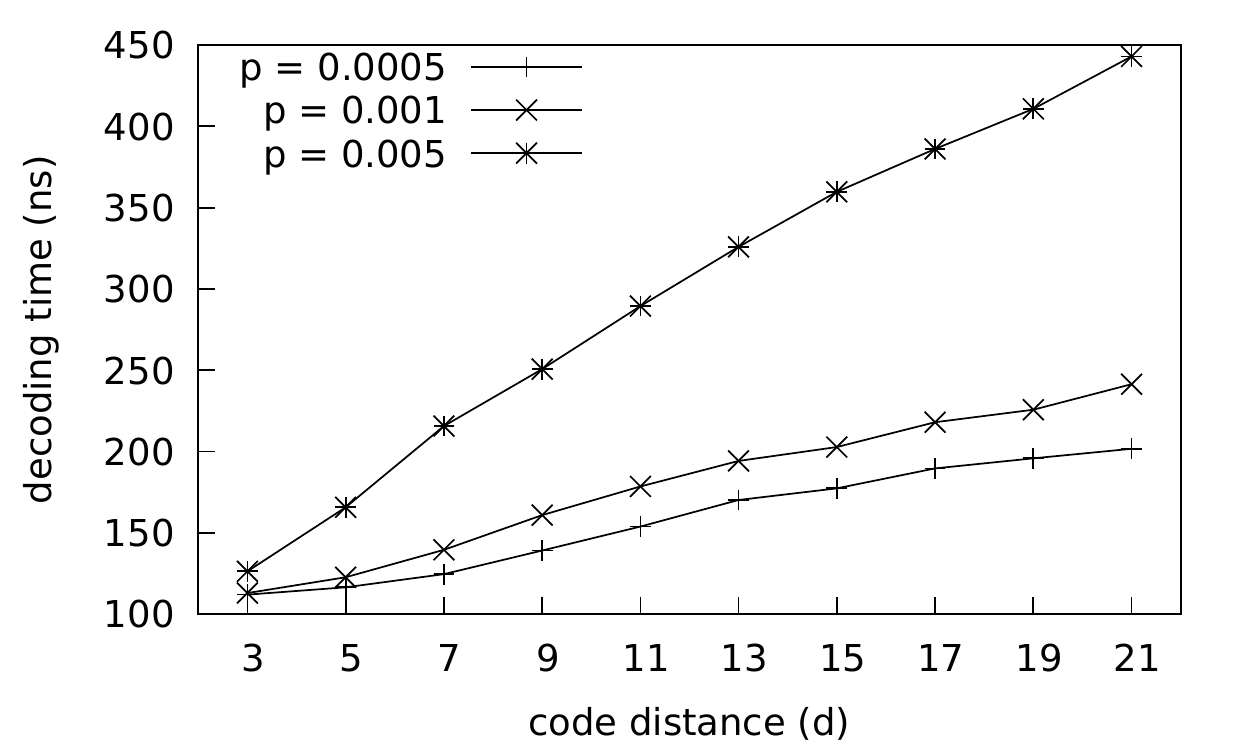}
        \label{fig:d_growth}
        \end{subfigure}
        \hspace{5ex}
        \begin{subfigure}{0.4\linewidth}
	    \centering
        \includegraphics[width=1.0\textwidth]{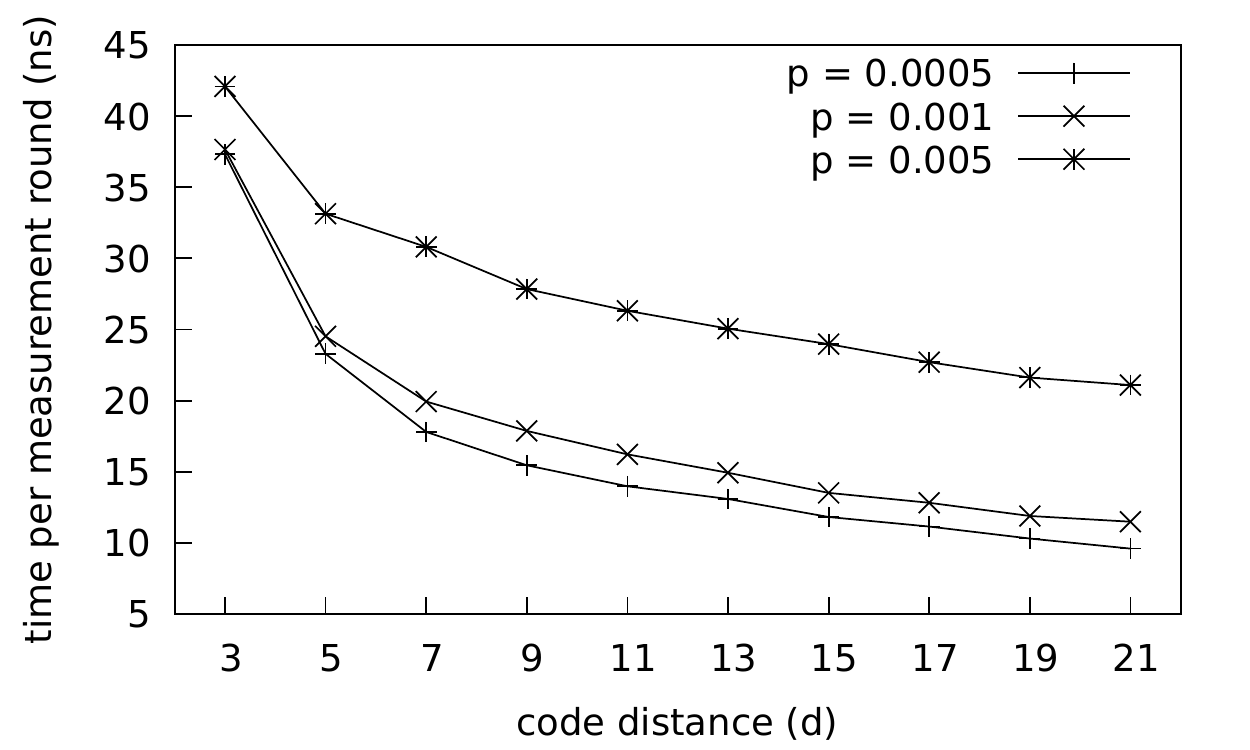}
        \label{fig:per_d_rounds}
        \end{subfigure}

	\caption{Average decoding time scales sub-linearly with $d$. We measure the average decoding time for 3 different noise levels. (Left) The average decoding time. (Right) The average decoding time per measurement round. The average time per measurement round reducing continuously justifies that our decoder is scalable for large surface codes. We show the distributions separately in \autoref{fig:d_variation}.}  \label{fig:DUF_decoding_time}
 \end{figure*}

To demonstrate the scalability of our algorithm with respect to the size of the surface code, we plot the average time for decoding against the size of the surface code. 
In \autoref{fig:DUF_decoding_time} (left) the y-axis shows the average decoding time in nanoseconds and the x-axis shows the distance ($d$) of the surface code.
We see that for all 3 physical error rates we tested, average decoding time grows sub-linearly with respect to the surface code size, which satisfies the scalability criteria to avoid an exponential backlog.
This implies that the average time to decode a measurement round reduces with increasing $d$ as shown in \autoref{fig:DUF_decoding_time} (right).



\paragraph{Distribution of decoding time}

\begin{figure*}[!t]
	    \centering
     \begin{subfigure}{0.32\linewidth}
         \centering
        \includegraphics[width=1.0\textwidth]{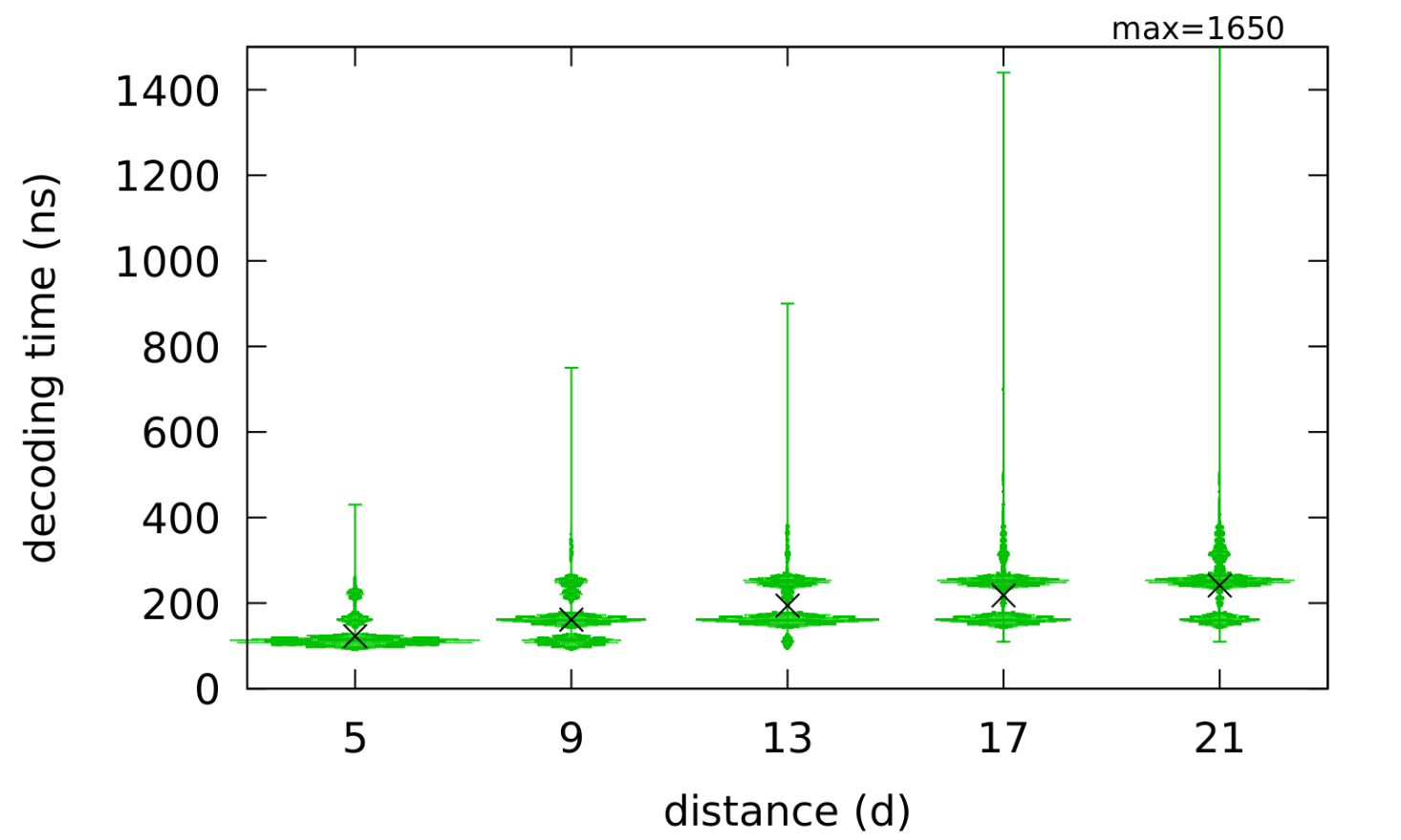}
	\caption{$T$'s distribution has a small mean \& a long tail}
 	\label{fig:d_variation}
\end{subfigure}
\hfill
\begin{subfigure}{0.32\linewidth}	    
\centering
     \includegraphics[width=1.0\textwidth]{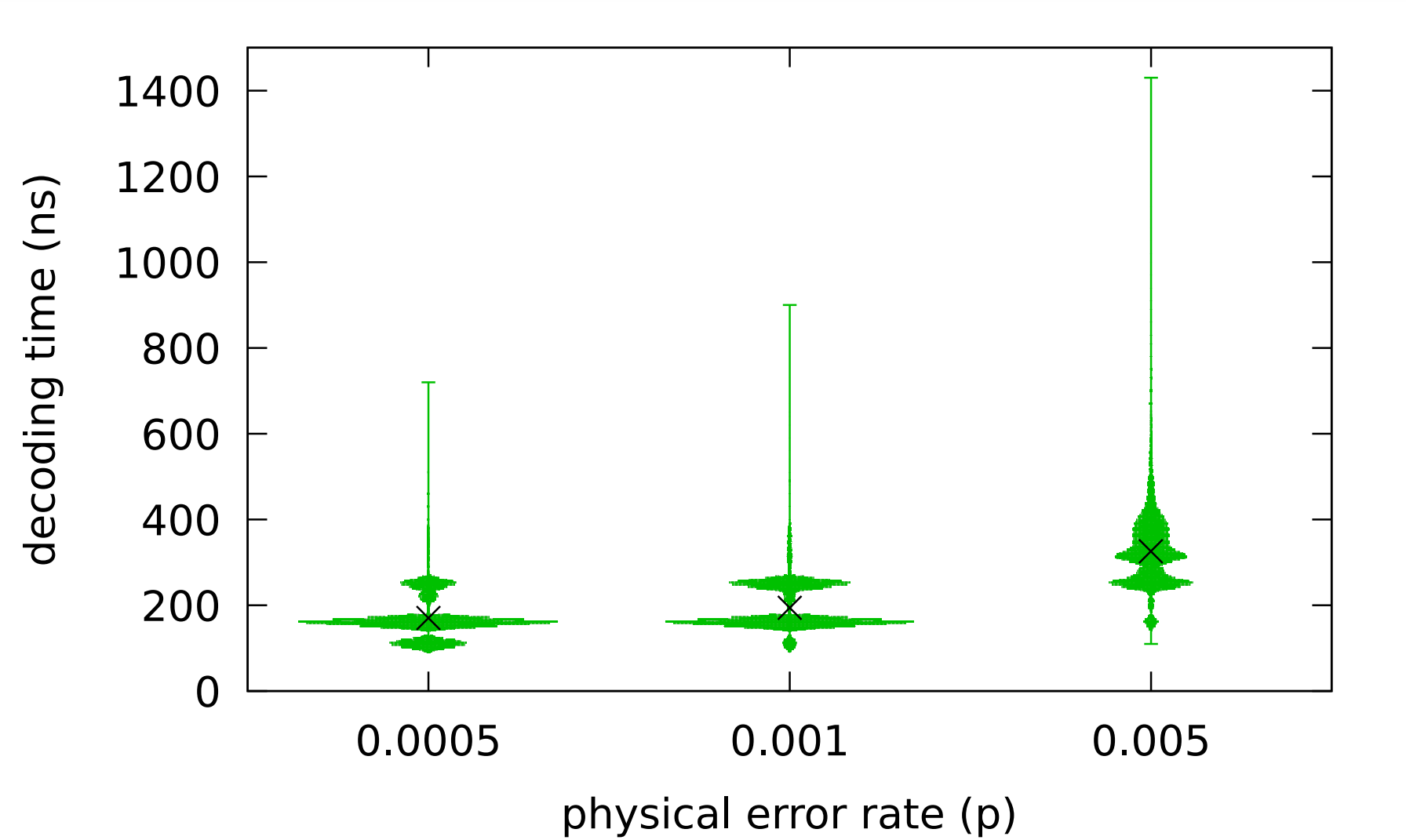}
	\caption{$T$ grows with the physical error rate.
 }
	\label{fig:noise_variation}
 \end{subfigure}
 \hfill
\begin{subfigure}{0.32\linewidth}	    
\centering
     \includegraphics[width=1.0\textwidth]{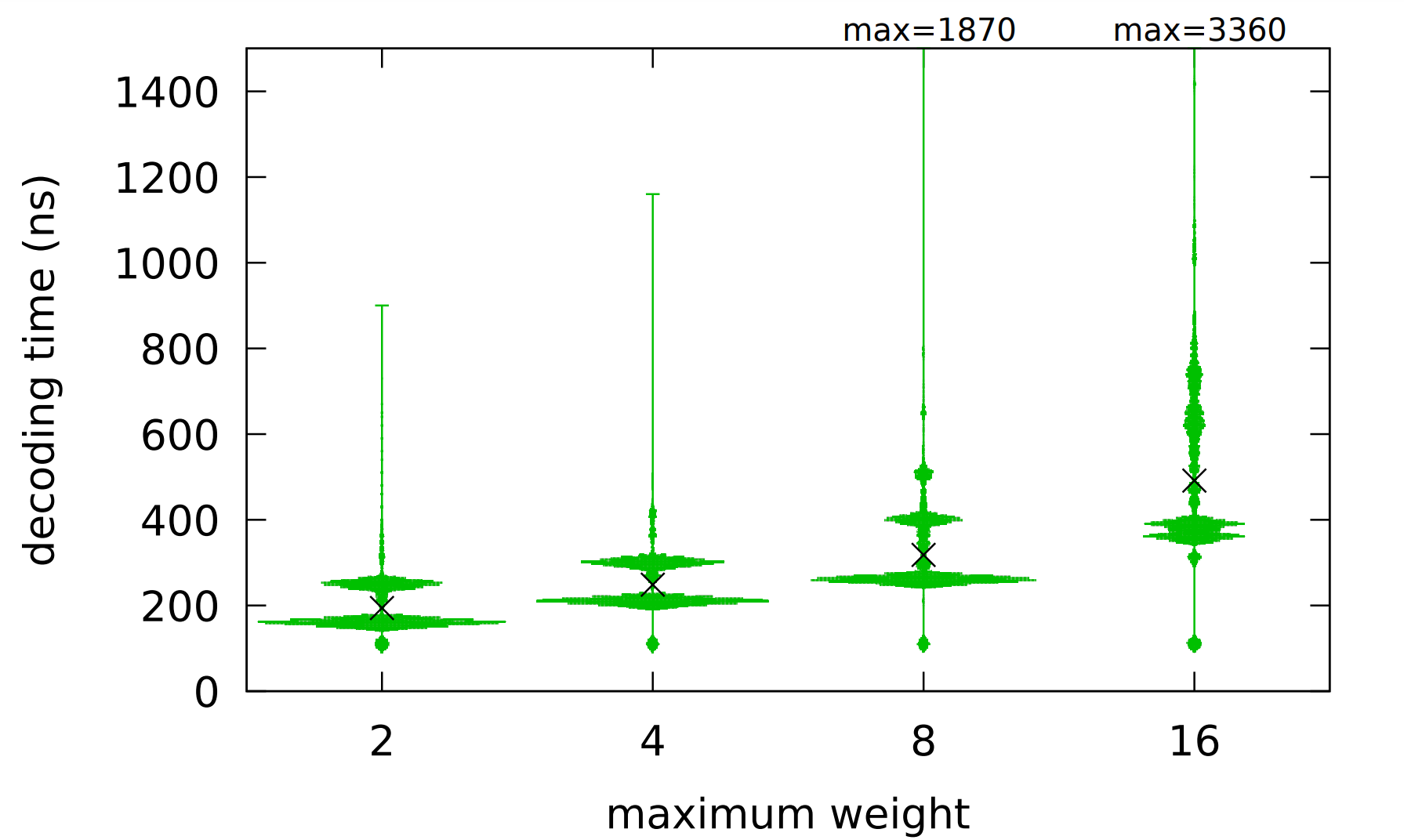}
	\caption{$T$ grows with the weight of the edges.
 }
	\label{fig:w_variation}
 \end{subfigure}
 \caption{Distribution of decoding time ($T$) with the mean marked with $\times$. Each distribution includes $10^6$ data points. By default $d=13$, $p=0.001$ and is unweighted }
  \end{figure*}

To understand the growth of decoding time with respect to the code distance, in \autoref{fig:d_variation} we plot the distribution of decoding time for different code distances.
The y-axis shows the decoding time and the x-axis shows the distance ($d$) of the surface code.
The average cycle count is indicated with $\times$. 

The key factor determining the decoding time is the number of iterations of growing and merging the distributed UF decoder requires. The peaks in the probability distribution for each distance in \autoref{fig:d_variation} correspond to the number of iterations. The variation around each peak is caused by the time required to sync $c\_id$ and calculate $odd$.  The number of iterations is related to the size of the largest cluster, which in turn correlates with the size of the longest error chain in the syndrome. As the size of the surface code increases, the probability of a longer error chain also increases, resulting in the probability distribution shifting to the right. 

Furthermore, as seen in \autoref{fig:d_variation}, the distribution for each surface code size is right-skewed. For example, for $d=13$, $90\%$ of trials required two iterations or fewer, which were completed within 250 ns. In the same test, $99.99\%$ of trials were completed within 370 ns. Only a very small number of error patterns require long decoding times, corresponding to syndromes with long error chains. Since such syndromes occur rarely and have poor decoding accuracy even if the decoding time is bounded, the impact on accuracy will be minimal. 

\paragraph{Effect of physical error rate}

To understand the effect of the physical error rate on decoding time, in \autoref{fig:noise_variation} we plot the distribution of latency for three different noise levels for $d=13$. The y-axis shows the latency and the x-axis shows the physical error rate.

As the noise level increases, the probability distribution of latency shifts to the right. This is caused by the increased probability of a longer error chain when the physical error rate increases, which in turn requires more iterations to decode. As a result, the average decoding time increases with the physical error rate.

\subsection{Non-identically distributed errors}

We next analyze the decoding process of a surface code with varying error probabilities for data and measurement qubits. While identically distributed errors are useful for evaluating the decoder's performance, practical implementation of surface codes may have different error probabilities for each qubit. To address this issue, each edge $i$ in the decoding graph is assigned a weight $w_i$ that ranges from 2 to $w_{max}$ and is proportional to $-log(p_i)$, where $p_i$ is the error probability corresponding to edge $i$. $w_{max}$ is a user-specified parameter indicating the resolution of error probabilities.

\textbf{Noise model :} 
We assign random error probabilities from a standard normal distribution with a mean of $0.001$ and a standard deviation of $0.0005$.

 ~\autoref{fig:w_variation} shows that the average latency increases as $w_{max}$ increases. When the errors have a higher resolution, more iterations are required for each cluster, leading to an increase in latency. For the unweighted graph with $d=13$, the average decoding time per round of 15 ns  increases to 38 ns when $w_{max}$ increases to 16. Notably, all of these values are significantly faster than the rate of measurement. As a result, decoding non-identically distributed errors can be performed in real-time using distributed UF on Helios.
 
\balance

\subsection{Comparison with related work}
\label{ssec:comparison}

Our empirical results as shown in \autoref{fig:d_variation} suggest that \name has a lower asymptotic complexity than any existing MWPM or UF implementation for which asymptotic complexities are available, e.g., ~\cite{fowler2014minimum,delfosse2017almost}. 
Indeed, the empirical results suggest that our decoder has a sub-linear time complexity: the decoding time per round decreases with the number of measurement rounds, which has never been achieved before.
This implies that \name can support arbitrarily large $d$ as the rate of decoding will always be faster than the rate of measurement. 

Das et al~\cite{das2022afs} calculate an average latency for their AFS decoder based on memory access cycles and assuming a design running at 4 GHz.
As the number of memory access cycles grows quadratically with $d$, the average decoding time per measurement round of AFS grows $O(d^2)$.
Similarly, Ueno et al~\cite{Ueno2021Qecool} estimate the decoding time of QECOOL from $d=5$ to $d=13$ based on SPICE-level simulations with a clock frequency of 5 GHz.
For the given range of $d$, the decoding time per measurement round increases quadratically with $d$.
In comparison, the decoding time of \name decreases per measurement round.

We should like to point out that AFS and QECOOL assume very high clock frequencies, which is key to their estimated low latency. 
For example, for $d=11$, AFS and QECOOL respectively report latencies of 42 ns and 8.32 ns per measurement round.
\name, in contrast, requires 16.2 ns per measurement round with a 100 MHz clock.


To the best of our knowledge, LILLIPUT ~\cite{das2021liliput} is the only hardware decoder in the literature that provides implementation-based results, for $d=5$.
The decoder has an average time of 21~ns per measurement round, which is slightly lower than that of \name for $d=5$, i.e., 24.5~ns.
However, as analyzed in \S\ref{sec:related}, LILLIPUT is not scalable for $d>5$.
Our work, in contrast, has successfully demonstrated the implementation of a $d=21$ surface code on a VCU129 FPGA with 11.5~ns per measurement round. 
The architecture of \name can potentially support larger $d$ using a larger FPGA, for example, $d=29$ for Xilinx VU19P~\cite{vu19fpga}, and even larger $d$ using a network of FPGAs.

Our decoder outperforms the two fastest software MWPM decoder, Sparse Blossom~\cite{higgott2023sparse} and Fusion Blossom~\cite{yueGithub}, by an order of magnitude. According to our evaluation, Sparse Blossom and Fusion Blossom take 160 ns and 295 ns per measurement round, respectively, for $d=13$ and $p=0.1\%$, using a single core of an M1 Max processor. In contrast, Helios achieves an average decoding time of 15 ns per measurement round under the same conditions, which is more than 60 times faster than the current state-of-the-art measurement rate~\cite{Chen2021Exponential}.

%% file: related_work.tex
\section{Related Work}
\label{sec:related}

There is a large body of literature on fast QEC decoding, e.g., \cite{battistel2023realtime, Terhal2015quantum, Gottesman2009Introduction, 2013LidarQuantumTopological}. The most related are solutions that leverage parallel compute resources.

Fowler~\cite{fowler2014minimum} describes a method for decoding at the rate of measurement ($O(d)$).
The proposed design divides the decoding graph among specialized hardware units arranged in a grid. 
Each unit contains a subset of vertices and can independently decode error chains contained within it. 
The design is based on the observation that large error patterns spanning multiple units are exponentially rare, so inter-unit communication is not frequently required.  
It, however, paradoxically assumes that the number of vertices per unit is ``sufficiently large'' and  a unit can find an MWPM for its vertices within half the measurement time on average. Not surprisingly, to date, no implementation or empirical data have been reported for this work.
Our approach uses vertex-level parallelism and leverages the same observation that communication between distant vertices is infrequent.

NISQ+\cite{Holmes2020Nisq} and QECOOL\cite{Ueno2021Qecool} parallelize computation at the ancilla level, where all vertices in the decoding graph representing measurements of one ancilla are handled by a single compute unit. 
This results in an increase in decoding time per measurement round as $d$ increases.
In contrast, we allocate a processing element per each vertex, which results in decreasing decoding time per measurement round with $d$ at the expense of the number of parallel units growing $O(d^3)$.
Furthermore, they both implement the same greedy decoding algorithm that has much lower accuracy than the UF decoder used in this work.
QECOOL has an accuracy that is approximately four orders of magnitude lower than that of a UF decoder~\cite{das2022afs} and NISQ+ ignores measurement errors further lowering its accuracy than QECOOL.

Skoric et al.~\cite{Skoric2022Parallel}, Tan et al.~\cite{tan2022scalable} and Wu~\cite{yueGithub} propose similar methods of using measurement round-level parallelism, in which a decoder waits for a large number of measurement rounds to be completed and then decodes multiple blocks of measurement rounds in parallel. 
By using sufficient parallel resources these methods can achieve a rate of decoding faster than the rate of measurement.
However, the latency of such approaches grows with the number of measurement rounds the decoder needs to batch to achieve a throughput equal to the rate of measurement.
In contrast, our approach exploits vertex-level parallelism and completes the decoding of every $d$ round of measurements with an average latency that grows sublinearly with $d$.

Pipelining can be considered a special form of using compute resources in parallel, i.e., in different pipeline stages. 
AFS \cite{das2022afs} is a UF decoder architected in three pipeline stages. 
The authors estimate the decoder will have a 42~ns latency for  $d=11$ surface code, which is $2.4$ times higher than what we report based on implementation and measurement.
The authors assume specialized hardware that is capable of running at 4~GHz and as a result, the decoding latency will be dominated by memory access.
However, no implementation or cycle-accurate simulation is known for this decoder.
Importantly, pipelining is limited in how much parallelism it can leverage: the number of pipeline stages.
In contrast, the parallelism  of our decoder grows along $d^3$, which enables us to achieve a sublinear average case latency.

LILLIPUT \cite{das2021liliput} is a three-stage look-up-table based decoder similar to AFS.  Look-up-table based decoders can achieve fast decoding but are not scalable beyond $d=5$ as the size of the look-up table grows $O(2^{d^3})$. For $d=7$ surface code with 7 measurement rounds, it would require a memory of $2^{168}$ Bytes, which is infeasible in any foreseeable future.

Sparse Blossom~\cite{higgott2023sparse}, a C++ MWPM implementation, decodes faster than the rate of measurement for $d=17$ on a single CPU core. 
However, its decoding time per round grows linearly with $d$ and increases to a few micro-seconds when the noise level increases, making it impractical for real-time decoding for higher noise levels and large surface codes. 
Fusion Blossom~\cite{yueGithub} takes a similar approach to Sparse Blossom and additionally parallelizes the computation at the measurement round level. 
By allocating 100 measurement rounds to each core on a 64-core processor, Fusion Blossom can decode up to $d=33$ faster than the measurement rate. 
However, both Fusion blossom and Sparse Blossom has a decoding time per round higher than that of \name by orders of magnitude, which limits their immediate use in quantum computing.
\lin{Check}

%% file: conclusion.tex
\section{Conclusion}
\label{sec:Conclusion}

We describe a distributed design for the Union Find decoder for quantum error-correcting surface codes, along with \name, a system architecture for its realization. 
Our FPGA-based implementation of \name demonstrates empirically that the average decoding time grows sub-linearly with the $d$. Using a VCU129 FPGA, \name decodes distance 21 surface codes at an average speed of 11.5 ns per measurement round, the fastest to the best of our knowledge.
\name is faster and more scalable than any previously reported surface code decoder implementations. 
Our results suggest that by leveraging parallel hardware resources, \name can avoid a growing backlog of measurements for arbitrarily large surface codes.

\section*{Acknowledgments}
This work was supported in part by Yale University and NSF MRI Award \#2216030.